\documentclass[12pt]{article}

\usepackage{graphicx}

\usepackage{amsmath}
\usepackage{amsxtra}
\usepackage{amstext}
\usepackage{amssymb}
\usepackage{latexsym}
\usepackage{dsfont}

\newcommand{\be}{\begin{equation}}
\newcommand{\ee}{\end{equation}}
\newcommand{\bn}{\begin{eqnarray}}
\newcommand{\en}{\end{eqnarray}}

\begin{document}

\title{\bf Neutrino mixing and masses in a left-right \\ model with mirror fermions }

\author{\small R. Gait\'an$^2$, A. Hern\'andez-Galeana$^1$, J. M.
Rivera-Rebolledo$^1$\\
\small and P. Fern\'andez de C\'ordoba$^3$\\
\small  "Interdisciplinary Modeling Group, InterTech."     \\
\small                                                 \\
\small  1. Departamento de F\'{\i}sica,\\
\small Escuela Superior de F\'{\i}sica y Matem\'{a}tica, I.P.N., \\
\small U.P. Adolfo L. Mateos, M\'{e}xico D.F., 07738, M\'{e}xico \\
\small                                                       \\
\small 2. Centro de Investigaciones Te\'oricas, FES, UNAM,\\
\small Apartado Postal 142, Cuatitl\'an-Izcalli, Estado de M\'exico,\\
\small C\'odigo postal 54700, M\'exico.}

\maketitle

\begin{abstract}
In the framework of a left-right model containing mirror fermions
with gauge group SU(3)$_{C} \otimes $SU(2)$_{L} \otimes $SU(2)$_{R}
\otimes $U(1)$_{Y^\prime}$, we estimate the neutrino masses, which are found
to be consistent with their experimental bounds and hierarchy. We
evaluate the decay rates of the Lepton
Flavor Violation (LFV) processes $\mu \rightarrow e \gamma$, $\tau
\rightarrow \mu \gamma$ and $\tau \rightarrow e\gamma$. We obtain upper limits for
the flavor-changing branching ratios in agreement with their present experimental
bounds. We also estimate
the decay rates of heavy Majorana neutrinos in the channels $N
\rightarrow W^{\pm} l^{\mp}$, $N \rightarrow Z \nu_{l}$ and $N
\rightarrow H \nu_{l}$, which are roughly  equal for large values of the heavy neutrino mass.
Starting from the most general Majorana neutrino mass matrix, the smallness of active
neutrino masses turns out from the interplay of the hierarchy of the involved
scales and the double application of seesaw mechanism. An appropriate parameterization
on the structure of the neutrino mass matrix imposing a symmetric mixing of electron
 neutrino  with muon and tau neutrinos leads to Tri-bimaximal mixing matrix for light neutrinos.
\end{abstract}


\pagebreak

\section{Introduction}

The evidences for neutrino oscillations obtained in experimental
results from atmospheric, solar, reactor and accelerator neutrinos
lead to conclude that the neutrinos have a mass different from zero.
The current neutrino experimental data (SuperKamiokande, SNO,
Kamland, K2K, GNO, CHOOZ) can be described by neutrino
oscillations via three neutrino mixings \cite{superkamiokande}.The
present data give the solar neutrino lepton mixing angle
$\tan^2\theta_{12} = 0.45 \pm 0.05$, the atmospheric angle $\sin^2
2\theta_{23} =1.02 \pm 0.04$ and $\sin^2 2\theta_{13} =0 \pm 0.05$
\cite{maltoni}. The complex phase has not yet been measured.

The experimental information on neutrino masses and mixing points out
new physics beyond the Standard Model (SM) of particle physics,
with a great activity on the consequences. Among the possible mechanisms of neutrino mass
generation, the most simple and attractive one is the seesaw
mechanism \cite{gellman,valle}, which explains the smallness of the
observed light neutrino masses through the exchange of superheavy
particles; an alternative explanation is given by extra dimensions beyond the usual three ones
\cite{aranda}. It has been suggested [ref.]that right-handed (RH)neutrinos
experience one or more of these
extra dimensions, such that they only spend part of their time in
our world, with apparently small masses. At the
present, it is not known whether neutrinos are Dirac or Majorana
fermions.

Models with heavy neutrinos of mass of order 1 $TeV$ can give rise
to significant light-heavy mixing and deviation from unitarity of
the Pontecorvo- Maki-Nakagawa-Sakata (PMNS) matrix
~\cite{langacker}. The nonunitarity nature of the neutrino mixing
matrix due to mixing with fields heavier than $\frac{M_{Z}}{2}$ can
manifest in tree level processes like $\pi \rightarrow \mu
\nu$, $Z \rightarrow \bar{\nu} \nu$, $W\rightarrow l \nu$ or in
charged lepton decays $\mu
 \rightarrow e \gamma$, $\tau
\rightarrow \mu \gamma$, etc. which are flavor violating and rare and proceed at one loop level
~\cite{langacker, nardi}. The $TeV$ scale seesaw models are
interesting because they can have signatures in the CERN Large
Hadron Collider (LHC) in the near future \cite{aguilar}.

Neutrinos also are important in astrophysics and cosmology
\cite{fukugita} and probably they contribute to hot dark matter in
the Universe and in its evolution.

Parity P violation was one of the greatest
discoveries of particle physics~\cite{wu}. Before this observation,
according to Fermi's hypothesis it was believed that weak
interactions have purely vectorial V or axial vectorial (V-A) parity
conserving Lorentz structure~\cite{fermi}. The theory of Lee and
Yang in 1956~\cite{yang} proposed a fermion current with V
and A structure. It is known that in the standard model (SM)
the electroweak interactions have a V-A form, with
only left-handed (LH) (ordinary) fermions coupling to the weak gauge
boson $W^{\pm}$. But one can include also mirror fermions \cite{maalampi}
with a $V+A$ coupling, such that P is conserved.
In this sense, the term "mirror fermion" is equivalent to
"vector-like fermion", where for a theory with gauge group $G$,
in a representation $R$ one has sets of LH and RH fermions.

In the literature a second meaning of that term  is used.
$G$ is extended to a $G \times G$ gauge theory, and for every multiplet $(R,1)$ a mirror
partner $(1,R)$ is added, such that there is no gauge invariant
mass term connecting the $LH$ and $RH$ multiplets ~\cite{okun1}.
Thus it is natural to consider the existence of mirror generations.

Masses of mirror particles arise from symmetry breaking; for mirror generation they may lye below one
$TeV$, and feasible to be discovered in Fermilab Tevatron Collider
and LHC.

A solution to the strong $CP$ problem has been proposed within a L-R symmetric context~\cite{barr}.
The electroweak group is extended to $SU(2)_{L} \otimes SU(2)_{R} \otimes U(1)$ including
mirror fermions. These fermions are conjugated to the ordinary ones
with respect to the gauge symmetry group such that a fermion
representation including both of them is real and the cancellation
of anomalies is automatic ~\cite{pati}.

In this paper we consider a L-R model with mirror fermions (LRMM) with gauge group
$G\equiv SU(3)_{C} \otimes SU(2)_{L} \otimes SU(2)_{R} \-\otimes
U(1)_{Y^\prime}$. We discuss in section 2 the
formalism of mixing between standard and new exotic fermions
In Sec. 3 we present the model and discuss the symmetry breaking
process with two scalar doublets.

In Sec. 4 we write the gauge invariant Yukawa couplings which after
spontaneous symmetry breaking give the most general Majorana
neutrino mass matrix. With a double application of the type I seesaw
approximation we estimate the light neutrino masses in terms of free
Yukawa couplings assuming textures for the light and mirror
matrices, obtaining consistent normal hierarchical values for masses
and a tribimaximal mixing for light neutrinos. We discuss in section
4 the mixing between standard and mirror fermions. In Sec. 5 we
include the radiative decays $\mu \rightarrow e \gamma$, $\tau
\rightarrow \mu \gamma$ and $\tau \rightarrow e\gamma$ and estimate
bounds for their branching ratios. Finally, we calculate such ratios
for the heavy Majorana neutrinos decays $N \rightarrow W^{+} l^{-}$,
$N \rightarrow Z \nu_{l}$ and $N \rightarrow H \nu_{l}$, getting a
smooth variation with the heavy neutrino mass, even when it is much
larger than any of the involved masses.


\section{Fermion mixing and flavor violation}
To consider the mixing of fermions, we shall follow Ref.
~\cite{langacker}, grouping all fermions of electric charge $q$
and helicity $a = L, R$ into $n_a + m_a$ vector column of $n_a$
ordinary (o) and $m_a$ exotic (e) gauge eigenstates, i.e.
$\psi^o_{a}= (\psi^o_{n_a}, \psi^o_{m_e})^{T}_{a}$. The ordinary fermions
include the SM ones, whereas the exotics include any new fermion
with sequential (mirror or singlet) properties beyond the SM.

The relation between the gauge eigenstates and the corresponding
light (l) and heavy (h) charged mass eigenstates $\psi_{a}= (\psi_l,
\psi_h)^{T}_{a}$, $a = L, R$ is given by the transformation \be
\psi^{0}_{a} = V_{a} \:\psi_{a} \;\;, \;\; a = L, R \ee

where

\be \label{chargedmixingmatrix} V_{a} = \left( \begin{array}{cc} A_a
& E_a \\ F_a & G_a
\end{array}
\right) \ee

\noindent In the Eq. (\ref{chargedmixingmatrix}), $A_a$ is a matrix
relating the ordinary weak states and the light-mass eigenstates,
while $G_a$ relates the exotic and heavy states. $E_a$
and $F_a$ describe the mixing between the two sectors.

From the unitary of $V$ \be \label{unitarityconditions} V_a V_a^{+}
= 1, a = L, R \ee it follows that the submatrix $A_a$ is not
unitary. The term $F_a^{+} F_a$, which is second order in the small
light-heavy fermion mixing, will induce flavor-changing transitions
in the light-light sector.

The vacuum expectation values (VEV) of the neutral scalars produce
the SM fermion mass terms, which together with the exotic mass and
mixing matrices lead to the mass matrix $M$ which takes the form

\be M = \left(
\begin{array}{cc}
K & \hat{\mu} \\
\mu & \hat{K}
\end{array}
\right) \ee

\noindent where $K$ denotes the $SM$ fermion mass matrix and
$\hat{K}$ corresponds to the fermion mass matrices associated with
the exotic sector, while $\mu$, $\hat{\mu}$ correspond to the mixing
terms between ordinary and exotic fermions.

The diagonal mass matrix $M_d$ can be obtained through a biunitary
rotation acting on the $L$ and $R$ sectors, namely \be M_d = V_L^{+}
M V_R = \left(
\begin{array}{cc}
m_l & 0 \\
0 & M_h
\end{array}
\right) \label{albi} \ee where $m_l$, $m_h$ denote the light and
heavy diagonal mass matrices, respectively.
The form of the mass matrix
will depend on the type of exotic fermion considered.


The scalar-fermion couplings within some specific Higgs sector are
not diagonal in general, and one can see that the couplings are not
diagonal in general; thus new phenomena associated with
flavor-changing neutral currents (FCNC) will be present in such
model.


\section{The Model}
In this and next sections we follow closely \cite{ceron}.
The LRMM formulation is based on the gauge group $SU(2)_{L}
\otimes SU(2)_{R} \otimes U(1)_{Y^{\prime}}$. In order to solve
different problems such as the hierarchy of quark and lepton
masses or the strong CP problem, different authors have enlarged
the fermion content to the form

\vspace{-5mm}
\begin{eqnarray}
\label{universal}
l^{0}_{i\, {L}} =
     \left( \begin{array}{c}
     \nu^{0}_i \\ {e^{0}_i}\end{array} \right)_{L}
   ,\ {e}^{0}_{i\,{R}}\
   ,\nu^{0}_{i\,{R}}, \ \ \ \ \ &;& \ \ \ \ \
 {\widehat l}^{0}_{i\, {R}} =
     \left( \begin{array}{c}
     {\widehat {\nu}^{0}_{i}} \\
     {{\widehat {e}}^{0}_{i}} \end{array} \right)_{R}
   ,\ {\widehat {e}}^{0}_{i\,{L}} \
   ,\ {\widehat{\nu}}^{0}_{i\,{L}},   \nonumber \\
Q^{0}_{i\, {L}} =
     \left( \begin{array}{c}
     {{u}^{0}_{i}}\\
     {{d}^{0}_{i}}
     \end{array} \right)_{L}
   ,\ {u}^{0}_{i\,{R}}\
   ,\ {d}^{0}_{i\,{R}}, \ \ \ \ \ &;& \ \ \ \ \
{\widehat Q}^{0}_{i\, {R}} =
     \left( \begin{array}{c}
          {\widehat {u}^{0}_{i}}\\
     {{\widehat {d}}^{0}_{i}}
     \end{array} \right)_{R}
   ,\ {\widehat {u}}^{0}_{i\,{L}}\
   ,\ {\widehat {d}}^{0}_{i\,{L}}\ ,
\label{Eq2.1}
\end{eqnarray}

\vspace{3mm}

\noindent where the index $i$ runs over the three fermion families and
the superscripts ${}^0$ denote gauge eigenstates. The quantum numbers
of these fermions under the gauge group $G$ defined above are given by

\vspace{-5mm} \bn l^{0}_{iL} \sim (1, 2, 1, -1)_{iL} \quad , \quad
\nu^{0}_{iR} \sim (1, 1,
1, 0)_{iR} \quad , \quad e^{0}_{iR} \sim (1, 1, 1, -2)_{iR} \nonumber \\
\widehat{\nu}^{0}_{iL} \sim (1, 1, 1, 0)_{iL} \quad , \quad
\widehat{e}^{0}_{iL} \sim (1, 1, 1, -2)_{iL} \quad , \quad
\widehat{l}^{0}_{iR}
\sim (1, 1, 2, -1)_{iR} \nonumber \\
u^{0}_{iR} \sim (3, 1, 1, \frac{4}{3})_{iR} \qquad , \qquad
d^{0}_{iR} \sim (3, 1, 1, \frac{2}{3})_{iR}   \nonumber \\
\widehat{u}^{0}_{iL} \sim (3, 1, 1, \frac{4}{3})_{iL} \qquad ,
\qquad \widehat{d}^{0}_{iL} \sim (3, 1, 1, \frac{2}{3})_{iL} \nonumber \\
Q^{0}_{iL} \sim (3, 2, 1, \frac{1}{3})_{iL} \qquad , \qquad
\widehat{Q}^{0}_{iR} \sim (3, 1, 2, \frac{1}{3})_{iR} \nonumber
\en

respectively, and the last entry corresponds to the hypercharge
($Y^{\prime}$) with the electric charge defined as $Q$ = T$_{3L}$
+ T$_{3R}$ + $\frac{Y^{\prime}}{2}$.

A model with gauge group $SU(2)_L \times SU(2)_R \times U(1)_V \times SU(3)_H$
and the fermion content (\ref{universal}) was originally suggested in Z. G. Berezhiani \cite{berezhiani1} as the "universal seesaw" model which generated masses of charged fermions as well as of the neutrinos. He also worked on a $SU(5) \times SU(3)_H$ model for extension to $SO(10)$ or Pati-Salam \cite{berezhiani2}, predicting for instance $m_{\nu_e} = O(10)$ eV. At low (electroweak scale) energies the model simulates the standard $SU(3)_C \times SU(2)_L \times U(1)_Y$ model, and FCNC are suppressed naturally.

\subsection{Symmetry breaking}
The "Spontaneous Symmetry Breaking" (SSB) is  achieved following the stages:

 \be \label{ssb} \mbox{G}\longrightarrow
\mbox{G}_{SM} \longrightarrow \mbox{SU(3)}_C \otimes
\mbox{U(1)}_Q\,  \ee

\noindent where G$_{SM}$ = SU(3)$_C \otimes $SU(2)$_L \otimes
$U(1)$_Y$ is the "Standard Model" group symmetry, and $\frac{Y}{2}$
= $T_{3R}$ + $\frac{Y^\prime}{2}$. The Higgs sector to induce
the SSB in Eq.(\ref{ssb}) involves two doublets of scalar fields:

\vspace{2mm}
\be \Phi = (1, 2, 1, 1) \qquad , \qquad \hat{\Phi} =
(1, 1, 2, 1) \ee

\vspace{2mm} \noindent where the entries correspond to the transformation
properties under the symmetries of the group $G$, with the "Vacuum
Expectation Values" (VEV's)

\be
<\Phi> = \frac{1}{\sqrt{2}}
           \left(
            \begin{array}{cr}
             0\\
             v
             \end{array}
             \right) \qquad , \qquad
<\hat{\Phi}> = \frac{1}{\sqrt{2}}
               \left(
               \begin{array}{cr}
               0\\
               \hat{v}
               \end{array}
               \right)\:.
\ee

\noindent The most general potential that develops this pattern of
VEV´s is

\vspace{-5mm} \be V = -(\mu \Phi^{\dag} \Phi + \hat{\mu}
\hat{\Phi^{\dag}} \hat{\Phi}) + \frac{\lambda_1}{2}[(\Phi^{\dag}
\Phi)^2 + (\hat{\Phi^{\dag}} \hat{\Phi})^2] + \lambda_2 (\Phi^{\dag}
\Phi) (\hat{\Phi^{\dag}} \hat{\Phi})]. \ee

In the last expression the terms with $\mu$, $\hat{\mu}$ are included so that the parity symmetry
(P)is broken softly, i. e., only through the dimension-two mass terms of Higgs potential.

\noindent The scalar Lagrangian for the model is written as

\be \label{covariant} {\cal L}_{sc} = (D_{\mu}\Phi)^{+}
(D^{\mu}\Phi) + (\hat{D}_{\mu}\hat{\Phi})^{+}
(\hat{D}^{\mu}\hat{\Phi}) \ee

\noindent where $D_\mu$ and $\hat{D}_{\mu}$ are the covariant
derivatives for the SM and the mirror parts,
respectively. The gauge interactions of quarks and leptons can be
obtained from the Lagrangian \vspace{-5mm}

\be
{\cal L}^{int} = \bar{\psi} i {\gamma}^{\mu} D_{\mu} \psi +
\bar{\hat{\psi}} i {\gamma}^{\mu} \hat{D}_{\mu} \hat{\psi}
\ee

The VEV's $v$ and $\hat{v}$ are related to the masses of the
charged gauge bosons $W$ and $\hat{W}$ by $M_{W}$ =
$\frac{1}{2} g_L v$ and $M_{\hat{W}}$ = $\frac{1}{2} g_R \hat{v}$,where
 $g_L$ and $g_R$ are the coupling constants of SU(2)$_L$ and
SU(2)$_R$, and $g_L$ = $g_R$ if we demand $L$-$R$ symmetry.

\pagebreak

\section{Generic Majorana neutrino mass matrix }

With the fields of fermions introduced in the model, we may write
the gauge invariant Yukawa couplings for the neutral sector\footnote{To simplify notation we drop the "0"
superscript }:

\bn h_{ij} \:\bar{\hat{\nu}}_{iL} \:\nu_{jR} + \lambda_{ij}
\:\bar{l}_{iL} \:\tilde{\Phi} \:\nu_{jR} + \eta_{ij} \:\bar{\hat{l}}_{iR}
\:\tilde{\hat{\Phi}} \:\hat{\nu}_{jL}  \nonumber         \\
                                       \nonumber   \\
+ \hat{M}_{ij} \:\bar{\hat{\nu}}_{iL} \:(\hat{\nu}_{jL})^c +
\sigma_{ij} \:\bar{l}_{iL} \:(\hat{\nu}_{jL})^c \;\tilde{\Phi}  \nonumber    \\
                                       \nonumber \\
+ \chi_{ij} \;\bar{\nu}_{iR} \;(\nu_{jR})^c + \pi_{ij}\:
\bar{\hat{l}}_{iR} \; (\nu_{jR})^c \:\tilde{\hat{\Phi}} + h.c. \en

\vspace{4mm}

\noindent where $i,j = 1,2,3$, $\tilde{\Phi}$= i$\sigma_2
\Phi^{*}$, $\tilde{\hat{\Phi}}$=i$\sigma_2 \hat{\Phi}^{*}$,
$h_{ij}$, $\hat{M}_{ij}$, $\chi_{ij}$ have dimensions of mass, and
$\sigma_{ij}$, $\eta_{ij}$, $\lambda_{ij}$ and $\pi_{ij}$ are
dimensionless Yukawa coupling constants. When $\Phi$
and $\hat{\Phi}$ acquire VEV's we get the neutrino mass terms

\bn h_{ij} \:\bar{\hat{\nu}}_{iL} \:\nu_{jR} + \frac{v}{\sqrt{2}}
\:\lambda_{ij} \:\bar{\nu}_{iL} \:\nu_{jR} + \frac{\hat{v}}{\sqrt{2}}
\:\eta_{ij} \:\bar{\hat{\nu}}_{iR} \:\hat{\nu}_{jL} \nonumber \\
+ \hat{M}_{ij} \:\bar{\hat{\nu}}_{iL} \:( \hat{\nu}_{jL} )^c +
\frac{v}{\sqrt{2}} \:\sigma_{ij}
\:\bar{\nu}_{iL} \:(\hat{\nu}_{jL})^c      \nonumber \\
+ \chi_{ij} \;\bar{\nu}_{iR} \;(\nu_{jR})^c + \frac{\hat{v}}{\sqrt{2}}
\:\pi_{ij} \;\bar{\hat{\nu}}_{iR} \;(\nu_{jR})^c + h.c. \en

\vspace{3mm} \noindent which are written in the generic
Majorana matrix form

\be \left( \overline{\Psi}_{\nu L} , \overline{\Psi^c}_{\nu L}
\right) \: \left(
\begin{array}{cc} M_L & M_D  \\ M_D^T & M_R \end{array}
\right) \: \left( \begin{array}{c} (\Psi_\nu^c)_R   \\
(\Psi_\nu)_R \end{array} \right) \label{generalmajoranamass} \ee

\vspace{3mm} \noindent where

\be (\Psi_\nu)_{L,R} = \left( \begin{array}{c} {\nu_i}   \\
\hat{\nu}_{i} \end{array} \right)_{L,R} \qquad , \qquad
(\Psi_\nu^c)_{L,R} = \left( \begin{array}{c} (\nu_i^c)   \\
(\hat{\nu}_i^c) \end{array} \right)_{L,R}  \ee

\vspace{4mm}

\be \label{majorana} M_L= \left( \begin{array}{cc}
0 & \frac{v}{\sqrt{2}} \: \sigma \\
                                 \\
\frac{v}{\sqrt{2}} \: \sigma^T & \hat{M}
\end{array}
\right) \qquad ,  \qquad M_R= \left(
\begin{array}{cc}
\chi & \frac{\hat{v}}{\sqrt{2}} \: \pi \\
                                       \\
\frac{\hat{v}}{\sqrt{2}} \: \pi^T & 0
\end{array}
\right) \;,\ee

\vspace{3mm}

\be \label{dirac} M_D= \left( \begin{array}{cc}
\frac{v}{\sqrt{2}} \: \lambda & 0 \\
                                  \\
h & \frac{\hat{v}}{\sqrt{2}} \: \eta
\end{array} \right)    \;,  \ee

\vspace{4mm}

\noindent with $h$, $\hat{M}$, $\chi$, $\sigma$, $\eta$, $\lambda$
and $\pi$ unknown matrices of $3 \times 3$ dimension. By assuming
the natural hierarchy $|(M_L)_{ij}| \ll |(M_D)_{ij}| \ll
|(M_R)_{ij}|$ for the mass terms, the mass matrix in
Eq.(\ref{generalmajoranamass}) can approximately be
diagonalized, yielding

\vspace{3mm}

\be \left( \overline{\Psi^\prime}_{\nu L} , \overline{
{\Psi^\prime}^c}_{\nu L} \right) \: \left(
\begin{array}{cc} M_\nu & 0  \\ 0 & M_R \end{array}
\right) \: \left( \begin{array}{c} ( {\Psi^\prime}_\nu^c)_R   \\
( {\Psi^\prime}_\nu)_R \end{array} \right) \;,\ee

\vspace{3mm} \noindent where, neglecting $
\mathcal{O}\:(M_D\:M_{R}^{-1}) $ terms, we may write in good
approximation\cite{sfk} ${\Psi^\prime}_{\nu L,R} \thickapprox \Psi_{\nu L,R}
$, and ${\Psi^{\prime \:c}}_{\nu L,R} \thickapprox \Psi^c_{\nu
L,R} $. The Majorana mass matrix for the left handed neutrinos may
be written in this seesaw approximation as

\vspace{3mm}

\be M_\nu \approx M_L - M_D\:M_R^{-1}\:M_D^T  \;.\ee

\vspace{3mm} \noindent We assume a scenario where the dominant contribution for the
active known neutrinos comes from the $M_L$ matrix having the
same structure of a Type I seesaw. Then in this scenario the eigenvalues
for the light neutrinos may be obtained by applying again the seesaw approximation,
that is:
\vspace{3mm}

\begin{equation}
M^{light} = - ( \frac{v}{\sqrt{2}}\:\sigma)\:{\hat{M}}^{- 1}\:( \frac{v}{\sqrt{2}}\:\sigma)^T
 \;.\label{lightneutrinos}  \end{equation}

\vspace{3mm} \noindent

Taking advantage of the fact that all $\sigma_{ij}$ and ${\hat{M}}_{ij}$ entries in
Eq.(\ref{lightneutrinos}) are free parameters, we propose the following
parameterizations for $\hat{M}$ and $M^{\text{light}}$ neutrino mass matrices:

\be M^{\text{light}}= \frac{Y^2 v^2}{2\:\hat{m}} \;
\begin{pmatrix} 1+b & b & b \\ b & 1+b+c & b-c \\
b & b-c & 1+b+c
\end{pmatrix} \qquad  , \qquad \hat{M}= \hat{m} \;\text{Diag}\:(Y_1,Y_2,Y_3) \;.
\label{TBMansatz}\ee

\vspace{3mm} \noindent where $Y$, $Y_1$, $Y_2$, $Y_3$, $b$, $c$ are dimensionless coupling constants and
$\hat{m}$ represents the mirror scale. This  parameterization for the light neutrinos
mass matrix imposes a symmetric mixing of electron neutrino with muon and tau neutrinos
in the first row and column of $( M^{\text{light}})_{ij}$, and the $2\times 2$ submatrix $i,j=2,3$
generate maximal mixing for muon and tau neutrinos. This structure for $M^{\text{light}}$  makes possible the diagonalization of light neutrinos by the so called "Tri-bimaximal mixing matrix" \cite{perkins}, i. e.

\begin{equation}
U_{\text{TB}}^T\:M^{light}\:V_{\text{TB}} = - U_{\text{TB}}^T\:( \frac{v}{\sqrt{2}\:\sigma})\:{\hat{M}}^{- 1} (\frac{v}{\sqrt{2}\:\sigma})^T\:U_{\text{TB}}
=Diag(m_1, m_2, m_3) \;,\end{equation}

\vspace{2mm} \noindent with

\be U_{TB} = \begin{pmatrix}
\frac{2}{\sqrt{6}} & \frac{1}{\sqrt{3}} & 0 \\
                                           \\
 - \frac{1}{\sqrt{6}} & \frac{1}{\sqrt{3}} &  - \frac{1}{\sqrt{2}} \\
                                             \\
  - \frac{1}{\sqrt{6}} & \frac{1}{\sqrt{3}} &  \frac{1}{\sqrt{2}}
\end{pmatrix} \ee

\vspace{2mm} \noindent and the light neutrino mass eigenvalues

\be (m_1, m_2, m_3)= \frac{Y^2 v^2}{2\:\hat{m}}\:(\:1,\: 1+3 b\:,\:1+2 c\:) \label{masslightneutrinos} \;.\ee

\vspace{2mm} \noindent The suppression by the mirror scale $\hat{m}$ in
Eq.(\ref{masslightneutrinos}) provides a natural explanation for the
smallness of neutrino masses. The allowed range of values for the square neutrino mass differences
reported in PDG \cite{PDG2010}:

\be  m_2^2 - m_1^2 \thickapprox 7.6 \times 10^{- 5} \;\text{eV}^2
\qquad , \qquad m_3^2 - m_2^2 \thickapprox 2.43 \times
10^{- 3} \;\text{eV}^2 \;,\ee

\vspace{3mm} \noindent with the input for normal hierarchy of the neutrino masses

\be ( \: m_1 \:,\: m_2 \:,\: m_3 \:)= (\:0.0865 \:,\: 0.0870 \:,\: .1 \:)\; \text{eV} \; ,
\label{inputneutrinomasses} \ee

\vspace{2mm} \noindent fix the parameter values as $b=0.00168$ and $c=0.07757$.
These neutrino masses are consistent with the bounds
$m_\nu < 2\:\text{eV}$ \cite{PDG2010}, and set the mass differences

\vspace{2mm}

\be  m_3^2 - m_1^2 \thickapprox 2.5 \times 10^{- 3} \;\text{eV}^2
\; . \ee

\vspace{2mm}

\noindent So, from Eqs.(\ref{masslightneutrinos}, \ref{inputneutrinomasses})

\vspace{2mm}

\be \frac{Y^2\:v^2}{2 \hat{m}} \approx 8.65 \times 10^{- 2}
\;\text{eV} \;.\label{RG} \ee

\vspace{4mm} \noindent Therefore, assuming $\hat{m}=m_{\hat{\nu}}=100\:\text{GeV}$ and $v=246\:\text{GeV}$ we obtain

\be Y \thickapprox 5.34 \times 10^{- 7} \ee

\vspace{3mm} \noindent
The matrix $M_L$ in Eq.(\ref{majorana}),  may be diagonalized by using a unitary
transformation

\vspace{3mm}

\be \label{mass1} U^{\dagger} \: M_L\:U = Diag
\left(m_1,m_2,m_3,\hat{m}_1, \hat{m}_2,\hat{m}_3\right)\:, \ee

\vspace{2mm}

\noindent where the mixing matrix $U$ compatible with our framework is written in
good approximation as

\vspace{2mm}

\be  \label{Nmixingmatrix}
U_{6 \times 6}  \thickapprox  \begin{pmatrix}
U_{TB} & \frac{v}{\sqrt{2}}\:\sigma\:{\hat{M}}^{-1} \\
                        \\
- ( \frac{v}{\sqrt{2}}\:\sigma\:{\hat{M}}^{-1} )^T & I_{3\times 3}
\end{pmatrix}  \;, \ee

\vspace{3mm} \noindent The particular numerical solution
congruent with the above scenario for the neutrino masses and mixing is

\vspace{2mm}

\be \frac{v}{\sqrt{2}}\;\sigma \approx 93041.9\: \text{eV} \: \begin{pmatrix}
 -1.2001 & 0.6355 & 1.2952 \\
 0.6355 & -1.2702 & 1.3006 \\
 1.2952 & 1.3006 & 0.5389
\end{pmatrix} \, ,\ee

\be \hat{M}= 100\:\text{GeV}\;\text{Diag} \left(\: 3.4918 \:,\: 3.2643 \:,\: 3.6043 \:\right) \; ,
 \ee

\noindent and

\be
\label{lightheavyNmixing}
\frac{v}{\sqrt{2}}\:\sigma \: {\hat{M}}^{- 1} \approx 9.3 \times 10^{- 7} \:\begin{pmatrix}
\begin{array}{ccc}
 -0.3437 & 0.1946 & 0.3593 \\
 0.1819 & -0.3891 & 0.3608 \\
 0.3709 & 0.3984 & 0.1495
\end{array}
\end{pmatrix} \ee

\vspace{4mm} \noindent for light $\nu$ - mirror mixing. Since the light-mirror
mixing is very small, the mixing matrix for light neutrinos behaves in good approximation
as the $U_{TB}$, Eq.(24). It is worth to mention here that in the limit of very small
light-mirror charged lepton mixing, $(F_L^\dag F_L )_{ij}\;,\;(E_L^\dag E_L )_{ij} \ll 1 $,
we may approach $U_{TB}$ as the usual $U_{PMNS}$ lepton mixing matrix for three generations.
Then, we obtain $(U_{PMNS})_{e2} \simeq\frac{1}{\sqrt{3}}$, $(U_{PMNS})_{e3} \simeq 0$, and
$(U_{PMNS})_{\mu 3} \simeq \frac{1}{\sqrt{2}}$, which give for the solar
and the atmospheric neutrino mixing angles $\theta_{12} \simeq  35.2^{0}$ and $\theta_{23} \simeq 45^{0}$,
with $\theta_{13} \simeq 0 $ in good agreement with current data,
although recent evidences \cite{newteta13bounds} show that $\theta_{13}$ may have a value different
from zero.

In earlier papers on the study of neutrinos and left-right symmetry
\cite{oliensis} appear similar representations of the fermions and
mass matrices as our in Eq.(\ref{dirac}), but these authors obtain
masses for the standard and mirror neutrinos some orders of
magnitude different from ours. On the other hand, the mass
generation in the LRMM here considered is
achieved with the scalar fields $\Phi$ and $\hat{\Phi}$,
Eqs.(3,4), transforming as doublets under $SU(2)_L$ and $SU(2)_R$,
respectively, with a mirror scale much lower than
$10^{12}$-$10^{13}$ $GeV'$s.


\section{Radiative decays}
In this section we analyze the lepton flavor violation processes
$\mu \rightarrow e \gamma$, $\tau \rightarrow \mu \gamma$ and $\tau
\rightarrow e \gamma$ arising in the model by the existence of gauge
invariant mixing terms between ordinary leptons and with the mirror counterparts.
The lower order contribution to theses decays mediated by the
neutral scalar fields comes from the Feynman diagrams where the
photon is radiated from an internal line. The corresponding
amplitude is proportional to the operator $\overline{u(p_2)}
\sigma^{\mu \nu} q_{\nu} \epsilon_{\mu} u(p_1)$, where $q = p_1 -
p_2$ and $\epsilon_{\mu}$ is the photon polarization \cite{lee}.

In the limit $m_e \ll m_{\mu} \ll m_{\tau}$ the rate decay is given
by

\be \Gamma(l_i \rightarrow l_j + \gamma) = \frac{\alpha}{512
\pi^{4}}(G_F m^2_{l_i})^2 \frac{m^5_{l_i}}{M_H^4}
|(\ln\frac{M_H^2}{m^2_{l_i}} - \frac{4}{3}) \epsilon_{i j}-
\sum_{k}{x_{\nu_k} V_{L,j k} V^{+}_{R,k i }}|^2 \ee

\noindent where $x_{\nu_k} \equiv \frac{m^2_{\nu_k}}{M^2_W}$,
$\epsilon_{i j} = |A_L^{+} A_R|_{ij}$ represents the flavor-changing
couplings, and the second term is the very small contribution from the light
neutrino propagating inside the loop.



\vspace{3mm} \noindent
In the limit $\alpha \ll 1$ and $M_{H} \ll M_{\hat{H}}$ the branching ratios are respectively
\be \label{muegamma}B_1(\mu \rightarrow e +
\gamma) = \frac{3 \alpha m^4_{\mu}}{8 M^4_H}
|(\ln\frac{M_H^2}{m^2_{\mu}} - \frac{4}{3}) \epsilon_{e \mu}-
\sum_{k}{x_{\nu_k} V_{L,e k} V^{+}_{R,k \mu}}|^2 \ee

\be B_2(\tau \rightarrow \mu + \gamma) = \frac{3 \alpha
m^4_{\tau}}{8 M^4_H} |(\ln\frac{M_H^2}{m^2_{\tau}} - \frac{4}{3})
\epsilon_{\mu \tau}- \sum_{k}{x_{\nu_k} V_{L,\mu k} V^{+}_{R,k
\tau}}|^2 \ee

\noindent and

\be \label{tauegamma} B_3(\tau \rightarrow e + \gamma) = \frac{3 \alpha m^4_{\tau}}{8
M^4_H} |(\ln\frac{M_H^2}{m^2_{\tau}} - \frac{4}{3}) \epsilon_{e
\tau}- \sum_{k}{x_{\nu_k} V_{L,e k} V^{+}_{R,k \tau }}|^2 \ee

By using the constraints $\epsilon_{i j} < 1\;,i \neq j$ for the parameters
in Eqs.(\ref{muegamma},\ref{tauegamma}), required by unitarity of $V$, see
Eqs.(\ref{chargedmixingmatrix},\ref{unitarityconditions}), one gets for the above
branching ratios:

\be B_1 < 2.2\times 10^{- 13} \quad , \quad  B_2 < 5\times 10^{- 9} \quad \text{and}
\quad  B_3 < 5\times 10^{- 9} \ee

\noindent which is congruent with the experimental bounds \cite{PDG2010} $B(\mu \rightarrow e + \gamma)< 1.2 \times
10^{-11}$, $B(\tau \rightarrow \mu + \gamma)< 4.4 \times
10^{-8}$ and $B(\tau \rightarrow e + \gamma)< 3.3 \times
10^{-8}$ PDG \cite{PDG2010}.

\section{Heavy Neutrino signals}

Possible new neutrinos can be detected in various ways in colliders.
If these neutrinos are heavy they will be unstable and may be detected directly in
their decay products.

Next generation of large colliders will probe Nature up to $TeV$
scales with high precision, probably discovering new heavy particles.
Thus, it will be a window to any new physics near the electroweak
scale which couples to the SM. Such colliders can be used to
produce new heavy neutrinos at an observable level to improve
present limits on their masses and mixings \cite{datta}. These
fermions with new interactions, like in the left-right models
\cite{robinett}, can be produced by gauge couplings suppressed by
small mixing angles. For the analysis of the heavy neutrinos signals
it is necessary to know their decay modes, which are different in the Dirac
and Majorana cases.

Heavy Majorana neutrino singlets can be produced
in the process \cite{delaguila}

\be q\bar{q^{\prime}} \rightarrow W^{*} \rightarrow l^{\pm} H
\ee

\noindent with $l = e, \mu, \tau$, which    cross sections depend on $M_N$ and the small
mixing $V_{l N}$. Heavy Majorana neutrino decays in the channels $N \rightarrow W^{\pm}
l^{\mp}$, $N \rightarrow Z \nu_{l}$ and $N \rightarrow H \nu_{l}$. The
partial widths for the $N$ decays are

\be
\label{maj1}
\Gamma(N \rightarrow
W^{+} l^{-}) = \Gamma(N \rightarrow
W^{-} l^{+}) =
\frac{e^2}{64 \pi s^2_{\theta_{w}}} |U_{l N}|^2
\frac{m^3_N}{M^2_W} (1 - \frac{M^2_W}{m^2_N}) (1 +
\frac{M^2_W}{m^2_N} - 2 \frac{M^4_W}{m^4_N})
\ee

\be
\label{maj2}
\Gamma(N \rightarrow Z \nu_{l})
= \frac{e^2}{64 \pi s^2_{\theta_{w}} c^2_{\theta_{w}}} |U_{l N}|^2 \frac{m^3_N}{M^2_Z} (1 - \frac{M^2_Z}{m^2_N})
(1 + \frac{M^2_Z}{m^2_N} - 2 \frac{M^4_Z}{m^4_N})
\ee

\be
\label{maj3}
\Gamma(N \rightarrow H \nu_{l})
=\frac{e^2}{64 \pi s^2_{\theta_{w}}} |U_{l N}|^2 \frac{m^3_N}{M^2_W} (1 - \frac{M^2_H}{m^2_N})^{2}
\ee

\noindent where $U_{l N}$ is the light-mirror neutrino mixing $\frac{v}{\sqrt{2}}\:\sigma \: {\hat{M}}^{- 1}$, Eq.(35). From
Eqs. (\ref{Nmixingmatrix},\ref{lightheavyNmixing}) the contributions come
from terms of the order $|V_{l N}| \lesssim 10^{-7}$. From these
expressions we can conclude that the total branching for each of
the four channels is independent of the heavy neutrino mixing, determined
only by $m_N$ and the gauge and Higgs boson masses.

Heavy neutrino signals are limited by the small mixing of the heavy neutrino required
by precision constraints \cite{victoria} and masses of order 100 $GeV$ are accessible at LHC.
For this mass range, SM backgrounds are larger and, since production cross sections are
relatively small, heavy neutrino singlets are rather difficult to observe.

\noindent The branching ratios for different values of $m_N$ reads as Table \ref{Brtable} ($M_H=130$ GeV);

 \begin{table}
 \begin{center}
 \begin{tabular}{ | c | c |c | c |}
 \hline $m_N(\text{GeV})$  & $B_{W^{\pm}}$  & $B_Z$  & $B_H$   \\
\hline   100 & 0.34  & 0.1  & 0.2   \\
 \hline  390 & 0.3 & 0.306 & 0.09  \\
 \hline  780 & 0.3 & 0.297  & 0.107 \\
 \hline  $\gg \:M_W, M_Z, M_H$ & 0.293 & 0.3  & 0.111 \\
 \hline
 \end{tabular}
 \end{center}
 \caption{ \label{Brtable} Branching ratios for different values of $m_N$  } \end{table}

\noindent and in all these cases $\sum B_i \approx 1 $. Here

\bn B_{W^{\pm}}=B_r (N \rightarrow W^{\pm} l^{\mp} ) \quad , \quad \quad B_Z=B_r (N \rightarrow Z \nu_l ) \quad ,\quad B_H=B_r (N \rightarrow H \nu_l )
\en

\noindent Table \ref{Brtable} shows that these decays are not so sensitive to
the heavy neutrino mass, such that for heavy neutrino signals it is not necessary
to have center of mass energies much larger than a hundred $GeV$.

Among the possible final states given by Eqs.(\ref{maj1}-\ref{maj3}),
only charged current decays give final states which may in principle be detected. For $m_N < M_W$ these two body decays are not possible and $N$ decays into three fermions,
mediated by off-shell bosons.

Other simple production processes like
\be
q \bar{q^{\prime}} \rightarrow Z^{*} \rightarrow \nu N
\ee

\be
g g \rightarrow H^{*} \rightarrow \nu N
\ee
give $l^{\pm}$ and $l^{+} l^{-}$ final states which are unobservable due to
the huge backgrounds. For the pair production
\be
q \bar{q} \rightarrow Z^{*} \rightarrow N N
\ee

\noindent the cross section is suppressed by $|V_{l N}|^4$, phase space and the $Z$
propagator, and is thus negligible.

Three signals are produced in the two charged current decay channels of
the heavy neutrino
\be
l^{+} N \rightarrow l^{+} l^{-} W^{+} \rightarrow l^{+} l^{-} l^{+} \bar{\nu}
\ee

\be
l^{+} N \rightarrow l^{+} l^{+} W^{-} \rightarrow l^{+} l^{+} l^{-} \nu
\ee
and small additional contributions from $\tau$ leptonic decays.

Heavy neutrino signals in the final state $l^{\pm} l^{\pm}$ are given in the lepton
number violating neutrino decay and subsequent hadronic $W$ decay, or leptonic
decay when the lepton is missed. LHC present energies are enough to discover heavy Majorana
neutrino with very small $V_{e\:N}$ \cite{pilaftsis}.

\section{Conclusions}
Here the LRMM with gauge group SU(3)$_{C} \otimes $SU(2)$_{L}
\otimes $SU(2)$_{R} \otimes $U(1)$_{Y^\prime}$ is applied in order to find closer values for neutrino masses fitted
to experimental data. We have worked with Majorana neutrinos, which
mass matrix was written in terms of blocks that stand for standard
and mirror mass terms. The large number of
parameters involved induces to make some
simplifications on the structure of the matrix. A double seesaw approach method is
used and diagonalization is
performed, and with the help of neutrino data we accommodate neutrino masses
with normal hierarchy of the order of
$(m_1, m_2, m_3) \approx (0.0865, 0.0870, 0.1 )\;\text{eV}$.
So, we have found a consistent smallness hierarchy
for the neutrino masses. With the LRMM we have also analyzed the
radiative decays $\mu \rightarrow e + \gamma$, $\tau \rightarrow e +
\gamma$ and $\tau \rightarrow \mu + \gamma$ for a Higgs mass of 130
$GeV$, obtaining bounds for the branching ratios congruent
with the experimental ones. Decay rates for heavy neutrinos $N$
were calculated for different channels, and we found that their BR are
nearly equal for $M_N \gg M_W , M_Z, M_H$ and also that they do not
change too much for other values of $M_N$. To find heavy
Majorana neutrinos one has only a few parameter dependence (for neutrino
singlets, the heavy neutrino mass and its mixing angle)and also the mass scale
could be accessible at the LHC.

\section{Acknowledgments}

The author R. Gait\'an wishes to thank to the "Sistema Nacional de
Investigadores" (SNI) in Mexico for partial support. and also
acknowledges support by PAPIIT project IN104208. A. Hernandez-Galeana is thankful for
partial support from the "Instituto Polit\'ecnico Nacional",
(Grants from EDI and COFAA) and "Sistema Nacional de
Investigadores" (SNI) in Mexico, and  J. M.
Rivera-Rebolledo wishes to thank to COFAA-IPN and the "Sistema Nacional de
Investigadores" (SNI) in Mexico for partial support.

\end{document}